\documentclass{article}
\usepackage{chicago}
\usepackage{latexsym}

\newcommand{\be}{\begin{equation}}
\newcommand{\ee}{\end{equation}}

\newcommand{\egc}{\mbox{e.\,g.\,}}

\newcommand{\dr}[1]{\ensuremath{\mathrm{d} #1\,}}
\newcommand{\mc}[1]{\ensuremath{\mathcal{#1}}}

\newcommand{\ket}[1]{\ensuremath{\left|  #1 \right\rangle}}
\newcommand{\bra}[1]{\ensuremath{\left\langle #1 \right|}}
\newcommand{\bk}[2]{\ensuremath{\left\langle #1 | #2 \right\rangle}}

\newcommand{\tpk}[2]{\ensuremath{\ket{#1}\!\otimes\!\ket{#2}}}

\newcommand{\matel}[3]{\ensuremath{\bra{#1} #2 \ket{#3}}}

\newcommand{\denop}{\ensuremath{\rho}}

\begin{document}

\title{Decoherence and its role in the modern measurement problem}
\author{David Wallace\thanks{Balliol College, Oxford University; \texttt{david.wallace@balliol.ox.ac.uk}}}
\maketitle

\begin{abstract}
Decoherence is widely felt to have \emph{something} to do with the quantum measurement problem, but getting clear on just \emph{what} is made difficult by the fact that the ``measurement problem'', as traditionally presented in foundational and philosophical discussions, has become somewhat disconnected from the conceptual problems posed by real physics. This, in turn, is because quantum mechanics as discussed in textbooks and in foundational discussions has become somewhat removed from scientific practice, especially where the analysis of measurement is concerned.

This paper has two goals: firstly (sections 1--2), to present an account of how quantum measurements are actually dealt with in modern physics (hint: it doesn't involve a collapse of the wavefunction) and to state the measurement problem from the perspective of that account; and secondly (sections 3--4), to clarify what role decoherence plays in modern measurement theory and what effect it has on the various strategies that have been proposed to solve the measurement problem.
\end{abstract}

\section{The myth of the ``conventional interpretation''}

Foundational discussions of quantum theory have a tendency to talk about the ``conventional'' or ``orthodox'' or ``standard'' interpretation of quantum mechanics. It's not generally very clear what's meant by this, but a rough stab might be that it consists of these two principles:
\begin{enumerate}
\item (the Measurement Algorithm) Observable quantities are represented by self-adjoint operators; the possible outcomes of a measurement of some observable are the eigenvalues of the associated operator; the probability of a given measurement outcome obtaining is given by the usual (Born) probability rule.
\item (the Projection Postulate) While in the absence of measurement a system evolves unitarily and deterministically, according to the Schr\"{o}dinger equation, when a measurement is made the system evolves stochastically, with its state vector being projected onto the eigensubspace corresponding to the actual measurement outcome. As such, the dynamics of quantum theory have a dual nature, with one evolution rule for non-measurement situations and one for measurement situations.
\end{enumerate}
Often (including by \citeN{Dirac1930}), the second is derived from the first by consideration of repeat measurements: presumably (so goes the reasoning) two measurements in succession had better give the same result, and this can only be guaranteed if the Projection Postulate holds.

It's generally held that this ``conventional interpretation\'' is profoundly unsatisfactory conceptually and philosophically, essentially because it treats ``measurement'' as a primitive term. Measurement, so the argument goes, is a physical process like any other, so (i) our theory should not contain assumptions which can be stated only by treating measurement as some primitive process, and (ii) whether a quantum system evolves according to one rule or another should not depend on whether we classify a physical process as a ``measurement'' or not.

My concern, however, isn't what makes the orthodox interpretation \emph{unsatisfactory}; it's what makes it \emph{orthodox}. To be sure: the orthodox interpretation is pretty much what we find in von Neumann's \citeyear{vonneumann} and Dirac's \citeyear{Dirac1930} original presentations of the subject. To be sure: it's pretty much what we find in most textbooks on quantum mechanics.\footnote{See, for instance, \citeN{cohentannoudji}, \citeN{Landau1977}, \citeN{Rae1992}, and \citeN{sakurai}; \citeN{ballentine} is an interesting exception.} And to be sure: it's pretty much what we find in most discussions of the quantum measurement problem. But it's not something which we actually find in use, much or at all, in mainstream applications of quantum theory. 

We can see this pretty straightforwardly in the case of the Projection Postulate. The postulate tells us how a quantum system evolves \emph{during measurement}, and this tells us immediately that it can only play a role in applications of quantum physics in situations where we want to analyse \emph{repeated measurements}. If all we care about is the outcome of single measurements, the Measurement Algorithm tells us all we need to know.

But if the point of the Projection Postulate is to analyse repeated measurements, there's an embarrassing problem: the Postulate tells us successive measurements of the same quantity always give the same result, and this is straightforwardly false in many --- perhaps most --- experimental contexts. (We can see this particularly dramatically in quantum optics: when we make a measurement of a photon, the photon is typically \emph{gone} at the end of the measurement.)\footnote{Bell and Nauenberg~\citeyear{bellmoral} pointed this out some while ago, and introduced a distinction between ``moral'' measurements which are compatible with the projection postulate, and ``immoral'' measurements, which aren't; see \citeN{homewhitakerzeno} for further discussion.}

The case of \emph{continuous} measurements (like those performed by a Geiger counter) also demonstrates that the Projection Postulate is unacceptable as an analysis of quantum dynamics during the measurement process. The only way to analyse continuous measurements via the Postulate would be to treat it as the limiting case of increasingly frequent measurements. But the quantum Zeno effect (\citeN{misrasudarshan}; see \citeN{homewhitakerzeno} for discussion) tells us that the result of any such limiting-case measurement is to freeze the system's evolution altogether, in flat contradiction of observed fact. (This was the original reason that Misra and Sudarshan, who did analyse continuous measurement in exactly this way, called the quantum Zeno effect a `paradox'.)

So if the only way the Projection Postulate could figure into the practice of physics is via the case of repeated measurements, and if it is patently unacceptable for that task, then there can be \emph{no} practical role for the Postulate in physics. And for what it's worth, crude sociological measures seem to back this up:  a search for ``projection postulate'', ``wave-function collapse'' and the like in the archives of Physical Review turns up only a few hundred references, nearly all of which turn out to be (a) foundational discussions, (b) discussions of proposed alternatives to quantum theory, or (c) theoretical quantum-computation discussions, where ``measurement'' does indeed get treated as primitive. (For comparison, searches for terms like ``state vector'' or ``Hilbert space'' or ``Schrodinger equation'' typically turn up several tens of thousands of references.) It's also notable that in \emph{relativistic} quantum mechanics, even defining the Projection Postulate is a delicate matter, because of relativity of simultaneity. One might then expect that textbooks on quantum field theory would need to address this point, but no such textbook of which I am aware does so; we can infer that the Projection Postulate is not much missed by practicing particle physicists or quantum field theorists.

However, I do not wish to suggest that we have \emph{no} effective way to analyse the process of repeated measurement. Actually, we can analyse it just fine (where here ``just fine'' means ``just fine if our goal is getting the empirically observed answer'', not necessarily ``just fine conceptually speaking''.) Doing so, though, requires us to drop the idea that measurement, as conceived in the Measurement Algorithm, is a primitive process.

This should have been obvious in any case. Treating ``measurement'' as a primitive term is not just philosophically unsatisfactory: it's a non-starter if you actually want to build and use a measurement device. And since we do actually build and use measurement devices, we can't just be treating measurement processes as primitive!

To expand: textbook and foundational discussions of quantum mechanics often give the impression that measurement devices are inexplicable black boxes which we find scattered across the desert, each one stamped with a self-adjoint operator and equipped with a funnel to insert quantum systems and a dial displaying the outcomes. But of course, real measurement devices are designed very carefully by experimental physicists. It's not a primitive feature of them that they measure whatever it is that they measure: it's a deliberate design feature. Perhaps at one time that design proceeded by handwaving appeals to quantum-classical correspondence, but in modern experimental physics the design of a quantum-mechanical measurement device relies extensively on quantum theory itself.

How can this be done? The answer, pretty much invariably, is that we apply quantum theory --- unitary quantum theory, with no Projection Postulate --- to  the measurement process itself, and then apply the Measurement Algorithm to the combined system of original-system-plus-measuring-device. This might seem like an infinite regress --- doesn't applying the Algorithm presuppose yet one more measurement? --- but in practice the regress can be ended when the system is macroscopically large, so that we can treat its macroscopic features --- say, its centre-of-mass position or momentum --- as straightforwardly observable properties. 

In the particular case of repeated measurements, a multiply-conducted measurement will need multiple readout displays. The measurement algorithm, applied to the combined original-system-plus-apparatus quantum system, delivers a determinate probability prediction over those displays. In some situations, it predicts that the displays are perfectly correlated; in \emph{these} situations, we could have got away with using the Projection Postulate. In most situations, the correlation will be either imperfect or absent altogether. Similarly, in the case of continuous measurements, analysing the actual measurement dynamics --- using \emph{unitary} quantum mechanics --- and then applying the Measurement Algorithm will give us the probability of a decay (say) being registered as a function of time (and will, in general, tell us that the decay rate is not especially sensitive to how closely the system is monitored, except in extreme cases).

We are left with an understanding of the measurement problem that is a better fit to contemporary scientific practice. There is no conventional \emph{interpretation} of quantum theory. What there is, is the \emph{formalism} of unitary quantum theory, shorn of any interpretation. We go from quantum theory to empirical predictions, not by understanding unitary quantum theory as a description of physical reality, but by applying a Measurement Algorithm, in principle at the level of microphysics but, when pushed, ultimately at the level of macroscopic, straightforwardly observable, facts. But that Algorithm (a) seems ill-defined and shot through with terms that don't belong in fundamental physics (``macroscopic'' and ``straightforwardly observable'' are only slightly more satisfactory than ``measurement'' as primitives), (b) relatedly, runs into the problem that even macroscopic observation processes are just more physical processes, and so shouldn't have special status, and (c) seems to block any attempt to understand quantum theory as a theory which tells us about Nature, and not just about what the dials and computer screens in our lab are going to do in any given situation.

From this point of view, it's also reasonably clear what would count as \emph{solving} the problem. We either need to find some way of \emph{understanding} unitary quantum theory so that it can give experimental predictions directly and not via the unsatisfactory Measurement Algorithm --- or we need to \emph{modify or supplement} unitary quantum theory so as to obtain a new theory which does not need the Measurement Algorithm.

\section{Two concepts of state space}

The account of the previous section may look unfamiliar. Normally (at least in foundational and philosophical discussions) the argument for a measurement problem is more direct: unitary quantum theory predicts macroscopic superpositions, which aren't observed, so unitary quantum theory must fail when applied to macroscopic systems (and presumably something like a Projection Postulate must be applied, albeit perhaps at the macroscopic level rather than for microscopic systems).

The basic assumption here is that Schr\"{o}dinger-cat states like
\be \label{catstate}
\alpha \ket{\mbox{cat alive}} +\beta \ket{\mbox{cat dead}}
\ee 
have to be understood as somehow describing a cat in an indefinite state of aliveness --- a cat that somehow is alive and dead at the same time. But why are we required to think this way?

In the philosophy literature, this reading of (\ref{catstate}) is often justified by appeal to the so-called
\begin{description}
\item[Eigenvalue-eigenvector link (E-E link):] A quantum system possesses a definite value of an observable if and only if its state is an eigenstate of the associated operator; if so, the definite possessed value is the eigenvalue associated with that eigenstate.
\end{description}
According to the E-E link, the state I wrote schematically as $\ket{\mbox{cat alive}}$ can be understood as a state of a live cat because it lies in some subspace of the cat's Hilbert space corresponding to the ``aliveness=1'' subspace of the ``aliveness operator''; similarly, the state $\ket{\mbox{cat dead}}$ is in the ``aliveness=0'' subspace and so is the state of a cat that is determinately not alive. The superposition of the two, lying in neither subspace, represents a cat that is neither alive nor dead.

This argument should not worry us, though, because the E-E link has no place in any serious discussion of the measurement problem. Like the Projection Postulate, it plays basically no role in physical practice (searching Physical Review's archives for the phrase and its variants gives precisely two hits); unlike the Projection Postulate, it seems to be purely an invention of philosophers, which does not appear in any quantum physics textbook of which I am aware. And this is not surprising, since in certain contexts it is fairly absurd, as can be seen easily by considering measurements of position. No realistic quantum wave-packet will remain in a bounded region of space for any non-zero length of time, unless kept there by infinite (and hence unphysical) potentials: in any physically realistic situation, some small section of the wavefunction will tunnel to arbitrarily large distances, arbitrarily quickly. So realistic quantum wavefunctions are spread right across space. Of course, macroscopic bodies can have wavefunctions that are, and that remain, extremely small outside some narrow region; wavefunctions like these are straightforwardly used in physics to represent localised bodies. But the fact remains that no such state lies in the eigenspace corresponding to any outcome of a position measurement, however coarse-grained.\footnote{In my view, failure to appreciate this point has caused much confusion in discussions of the so-called ``problem of tails'' in the GRW dynamical-collapse theory (cf \citeN{wallaceashgate} and references therein): the ``problem'', as generally discussed, has little to do with the GRW theory per se, but is a general feature of quantum mechanics.} So according to the E-E link, all realistic quantum systems have completely indefinite positions. If the E-E link were an inextricable part of quantum physics, this would count as a \emph{reductio ad absurdum} of quantum theory; as it is, it just counts as a \emph{reductio ad absurdum} of the E-E link.

So much for the E-E link. But even if \emph{that} argument for the unacceptable weirdness of Schr\"{o}dinger-cat states does not go through, there might perfectly well be \emph{other} arguments to the same conclusion. To get more perspective on states like (\ref{catstate}), it helps to think a bit about the concept of a state space in general.

The paradigm example of a state space is classical phase-space (I could tell essentially the same story with configuration space). States --- points in phase space --- represent the physical features of the system in question; different points in phase space represent physically distinct systems.  But this is not the only kind of state space definable in classical physics: the space of \emph{probability functions} on phase space can also be treated as a state space (call it distribution space). Mathematically, phase space and distribution space are just two sets of abstract mathematical objects with deterministic evolution equations: Hamilton's equations in the one case, Liouville's equations in the other. The central difference is conceptual: distinct points in distribution space do not represent systems in physically distinct states, but distinct probability distributions over physical states. In other words, there are two conceptions of state space available: physical, or probabilistic.\footnote{An alternative terminology due to \citeN{spekkensepistemic} calls the physical conception \emph{ontic} and the probabilistic conception \emph{epistemic}. (In the case of the quantum state, this becomes `$\psi$-ontic' and `$\psi$'-epistemic --- hence Chris Fuch's wonderfully cruel term (in conversation) for those who take the ontic view: ``$\psi$-ontologists''.) This terminology, though, suggests a particular reading of probability as a quantification of our ignorance, which I want to avoid being committed to.} 

One way to understand the difficulty of making sense of quantum states is that we seem to shift between these two conceptions. When dealing with states like \ref{catstate} in practice, we treat them as probabilistic, conveying nothing more or less than ``the cat is either alive or dead, and it has chance $|\alpha|^2$ of being in the first state, and $|\beta|^2$ of being in the second''. On this reading of the quantum state, there is nothing at all mysterious about (\ref{catstate}), and no need to invoke the Projection Postulate. Indeed, something like the Projection Postulate emerges spontaneously from the ordinary calculus of probability: it is a commonplace that probability distributions evolve \emph{both} under the ordinary dynamics of the system, \emph{and} via probabilistic updating when we acquire more information. From that perspective, the ``collapse'' of (\ref{catstate}) onto either $\ket{\mbox{alive cat}}$ or $\ket{\mbox{dead cat}}$ upon measurement is no more mysterious than the collapse of the classical probability distribution ``Heads with probability 1/2, tails with probability 1/2'' onto ``Heads'' or ``Tails'' upon measurement.

When dealing with the quantum states of microscopic systems, however, this straightforward probabilistic reading of the quantum state breaks down. Partly this is because the democracy of bases in Hilbert space allow a multitude of ways of expressing a state as a superposition of eigenstates: is the spin state
\be \label{spinstate}
\alpha\ket{+_z}+\beta\ket{-_z}
\ee 
to be interpreted as a probabilistic mixture of a particle with $z$-spin up and a particle with $z$-spin down, or as a different probabilistic mixture of $x$-spin up and $x$-spin down? (Or, somehow, mysteriously, as both at once?) But more crucially, the probabilistic reading simply fails to make sense of interference phenomena. Suppose that we set up a Mach-Zender interferometer,  in which a beam of (say) photons is split into left(L) and right(R) beams by a half-silvered mirror and then re-interfered by means of another such mirror, and the resultant beams are fed into two detectors, $A$ and $B$. We can easily establish, by blocking the R beam, that if the photon is originally in the L beam, it has a 50\% chance of ending up at detector $A$ after the second half-silvered mirror, and a 50\% chance of ending up at detector $B$. We can easily establish that the same is true if it is originally in the R beam. But in that case, which beam it is in makes no difference to the outcome probabilities, so \emph{any} probabilistic mixture of the two possibilities should lead to a 50\% chance of each of the two results. And of course, this is not what happens: depending on the relative phase of the two beams, any outcome from $100\% A$ to $100\% B$ is possible.\footnote{Nor will it do to say ``quantum probabilities just behave differently from classical ones''. The point of the argument is that we cannot explain the observed phenomena allowing the two ``possibilities'' to influence one another. But if something that can have a dynamical effect on a real thing doesn't thereby count as real, we lose our grip on reality. (Cf discussions in \citeN[chapter 4]{deutschfabric}, \citeN[chapter 10]{wallacebook}, or \citeN{anandanbrown}.) The bottom line is that a microscopic superposition cannot be understood as merely probabilistic: both terms in the superposition represent physical features of the world. But if the same is true for the Schr\"{o}dinger-cat state, the paradoxical nature of that state is not alleviated by the supposedly ``probabilistic'' reading.}

The two objections are related, of course. Interference phenomena can occur in quantum mechanics precisely because amplitudes have phases as well as magnitude. The probabilistic reading of a quantum state like (\ref{catstate}) or (\ref{spinstate}) interprets the magnitudes of the amplitudes $\alpha$ and $\beta$ as probabilities, but provides no interpretation of their phases. On this reading, replacing $\alpha$ with $\alpha \exp(-i\theta)$ should have no physical significance --- but of course, doing so not only has ``significance'', it has empirically detectable consequences.

This provides an alternative way to state the measurement problem:
\begin{quote}
We cannot consistently understand the state space of quantum theory \emph{either} as a space of physical states, \emph{or} as a space of probability distributions. Instead, we have to use one interpretation for microscopic physics and another for macroscopic physics. Furthermore, both the point at which we have to transition between the physical and probabilistic interpretation, and the basis with respect to which the probabilistic interpretation is to be specified, are defined only in an approximate, rough-and-ready way, which seems to make essential use of terms 
like ``macroscopic'' which have no place in a fundamental physical theory.
\end{quote}

It also provides a ready way to understand the various strategies that have been proposed to resolve the measurement problem. Firstly, there are the attempts to \emph{solve} the problem, by replacing quantum theory by a new theory which can be understood in a consistent way. There are two basic strategies:
\begin{description}
\item[Hidden-variable theories]
These theories hold on to the dynamics of unitary quantum theory, but remove the probabilistic interpretation of the quantum state. Instead, some new physical entities are introduced, which are dynamically influenced by the quantum state but not vice versa. Probability is then introduced via a probability distribution over these ``hidden variables''. The paradigm example is the de Broglie-Bohm pilot-wave theory, also called Bohmian mechanics,\footnote{See \citeN{cushingbohmbook} and references therein.} in which the quantum state is supplemented by a collection of point particles whose collective evolution is determined by the quantum state. (So-called `modal interpretations'\footnote{See \citeN{dieksvermaas} and references therein.}  also fit this category.)
\item[Dynamical-collapse theories]
These theories\footnote{Examples include the GRW theory (\citeN{grw}; see \citeN{bassighirardireview} for discussion) and proposals due to Penrose \citeyear{penroseenm}; cf also Philip Stamp's discussion in this volume).} try to make sense of the transition from a physical to a probabilistic reading of the quantum state not as an interpretative shift but as an objective physical process. This is generally done by introducing some version of the Projection Postulate that objectively removes macroscopic superpositions, and does so in a probabilistic manner. According to dynamical-collapse theories, the quantum state is always to be interpreted physically and its stochastic evolution deviates from the deterministic Schr\"{o}dinger equation in certain circumstances --- but it does so in such a way that the resultant probabilities are very close to those defined by the probabilistic reading of unitary quantum mechanics.
\end{description}

Secondly, there are the attempts to \emph{dissolve} the problem, by finding a consistent interpretation of unitary quantum theory. There are only two real options here:
\begin{description}
\item[The quantum state is always probabilistic]
This approach tries to interpret quantum state space as systematically like the space of classical probability distributions. This has proved extremely difficult, though, for essentially the reasons given above: interference phenomena don't seem to be understandable as probabilistic phenomena. More rigorously, the theoretical results of Bell~\citeyear{bell1966}, Kochen and Specker~\citeyear{kochenspecker}, and Gleason~\citeyear{gleason}, make it clear that any such strategy will have the apparently pathological feature of \emph{contextuality}.\footnote{For detailed discussion, see, \egc, \citeN{redheadbook}.} For this reason, most\footnote{Not all: \citeN{spekkens} is a counter-example.} adherents of this strategy have given up on the idea of interpreting the ``probabilities'' as probabilities of any microphysical even, and just read them directly as probabilities of measurement outcomes, treated as primitive. That is, they fall back on an instrumentalist reading of quantum theory. (For a clearly articulated example, see \citeN{fuchsperes}.) 
\item[The quantum state is always physical] This approach, most famously associated with the name of Hugh Everett~\citeyear{everett}, tries to interpret quantum state space as systematically like classical phase space. This deals straightforwardly with the problem of microscopic interference, but struggles with the problem of macroscopic superpositions. If Schr\"{o}dinger-cat states like \ref{catstate} are to be understood as physical states, what physical situation are they representing? The only coherent answer seems to be: a situation with two independently-evolving cats, one alive and one dead. But what justifies this interpretation, and where do the probabilities enter the theory, if they are not to be added directly through either a probabilistic reading of the state, or through additional hidden variables, or through stochastic dynamics?\footnote{For detailed discussion of the Everett interpretation, see Saunders \emph{et al}\citeyear{saundersetal} or \citeN{wallacebook}.}
\end{description}

However, once again it will be useful not to consider the conceptual question of how to \emph{make sense} of a state space that seems sometimes to be a space of physical states and sometimes a space of probability distributions, but to ask, more practically, how it is that we sometimes \emph{get away} with treating the state space as a space of probability distributions, given that fundamentally that interpretation does not seem consistent.

\section{The role of decoherence}

Let us consider more carefully just when the probabilistic interpretation can be consistently applied. Recall that it is \emph{interference} that blocks a probabilistic interpretation at the microscopic scale; if a probabilistic interpretation can be applied at any scale, then, it must be because interference phenomena can be neglected.

This, of course, is precisely the question which decoherence theory is designed to answer. In the following, I apply the environment-induced decoherence framework\footnote{See Joos \emph{et al}~\citeyear{joosetal} and references therein for a review of this framework.} used by Zeh~\citeyear{zeh93}, Zurek~\citeyear{zurek91} \emph{et al}; essentially the same conclusions, though, can be derived in the decoherent-histories framework (See, \egc,\citeN{gellmannhartle93}, \citeN{Griffiths1993}, \citeN{halliwellhydrodynamic}; see also my discussion in chapter 3 of \citeN{wallacebook}).

Firstly: suppose that the physical system we are considering has a Hilbert space \mc{H} which can be decomposed as 
\be 
\mc{H}=\mc{H}_M\otimes \mc{H}_E,
\ee
where $\mc{H}_M$ is the Hilbert space of the degrees of freedom that interest us --- say, the centre-of-mass degrees of freedom of a macroscopic object, or the low-momentum vibrational modes of an extended solid --- and $\mc{H}_E$ is the Hilbert space of all the other degrees of freedom that interact with those that interest us. $\mc{H}_M$ is normally said to represent the ``system'', and  $\mc{H}_E$  the ``environmental'', degrees of freedom, but it's important to recognise that the ``environment'' need not be spatially external: for the extended solid, for instance, $\mc{H}_E$ might be the Hilbert space of vibrational modes above some momentum cutoff. 

Suppose also that there is some basis $\{\ket{z}\}$, of $\mc{H}_M$, labelled by some (discrete or continuous) variable $z$, with the following property: the state
\be 
\left(\int \dr{z} \alpha(z)\ket{z}\right)\otimes \ket{\psi},
\ee 
(where the integral sign schematically denotes summation or integration over $z$, as appropriate) evolves rapidly, for fairly generic $\ket{\psi}\in \mc{H}_E$, to
\be \label{zstate}
\int \dr{z} \alpha(z)\tpk{z}{\psi(z)},
\ee 
where (i) `rapidly' means `rapidly as compared to the dynamical timescales of the system's own dynamics, and (ii) $\bk{\psi(z)}{\psi(z')}\simeq 0$ unless $z \simeq z'$. Effectively, dynamics like this consists of the environment `measuring' the system state. The classic example has $\{\ket{z}\}$ as a wave-packet basis, reasonably localised in both centre-of-mass position and momentum. Since the system-environment interactions are local, two wave-packets with significantly distinct locations will cause significantly different evolution of the environment. And two wave-packets with significantly different momentum will swiftly evolve into two wave-packets with significantly different position. (Handling this case carefully requires us to deal with the fact that wavepackets form an overcomplete basis; I'll ignore this issue for simplicity.)

If this occurs, we will say that the system is \emph{decohered} by the environment, with respect to the $\{\ket{z}\}$ basis.


Now suppose we want to apply the probability interpretation with respect to the $\{\ket{z}\}$ basis: that is, we want to treat a state like (\ref{zstate}) as having probability\footnote{Or probability density, if $z$ is a continuous variable.} $|\alpha(z)|^2$ of having whatever physical property (call it $Z$) that $z$ is supposed to represent (some given centre-of-mass position and/or momentum, say). If the density operator of the system at time $t$ is $\denop(t)$, then we have 
\be 
\mathrm{\Pr}(Z=z,t)=\matel{z}{\denop(t)}{z}
\ee
where $\mathrm{\Pr}(Z=z,t)$ is the probability at time $t$ of the system having $Z=z$.

For this really to be interpretable as a \emph{probability}, though, its evolution needs to be free from interference effects: that is, it needs to evolve \emph{like a probability function}. So: let $\Pr(Z=z',t';Z=z,t)$ be the probability\footnote{Again, this should be interpreted as a probability density if $z$ is continuous.} that the system has $Z=z'$ at time $t'$, \emph{given} that it has $Z=z$ at time $t$. Then the standard laws of probability tell us that
\be 
\Pr(Z=z',t')=\int \dr{z}\Pr (Z=z',t'|Z=z,t)\times \Pr(Z=z,t).
\ee 
The dynamics will have this form iff $\Pr(Z=z',t')$ is a linear functional of $\Pr(Z=z,t)$; that is, if $\matel{z}{\denop(t')}{z}$ (considered as a function of $z$) is a linear functional of $\matel{z}{\denop(t)}{z}$. 

In general, this will not be the case. Assuming we can write down autonomous dynamics for $\denop$ in the first place, then of course the overall linearity of unitary dynamics implies that  $\denop(t')$ is a linear functional of $\denop(t)$. But in general, $\matel{z}{\denop(t')}{z}$ depends not only on diagonal terms of $\denop(t)$ like $\matel{z}{\denop(t)}{z}$, but on off-diagonal terms like $\matel{z}{\denop(t)}{w}$. 

However, in the case we are considering, the interaction with the environment guarantees that off-diagonal terms in $\denop$ are suppressed, and suppressed on timescales much quicker than those which characterise the dynamics of the system itself. In \emph{this} situation, then, the effective evolution equation for $\denop$ reduces to an evolution equation specifically for the diagonal elements of $\denop$ in the $\{\ket{z}\}$ basis. 

Let's sum up. If the total system we are studying can be decomposed into ``system'' and ``environment'' degrees of freedom, such that for some basis $\{\ket{z}\}$ of the system, the system is decohered by the basis with respect to that basis, then we can consistently treat the total system as a probabilistic mixture of states like $\tpk{z}{\psi(z)}$. 

Furthermore, we have excellent reason --- from general physical arguments, from specific models, and increasingly from experiment --- to think that the macroscopic degrees of freedom of a system are decohered by the residual degrees of freedom with respect to a wave-packet basis for those macroscopic degrees of freedom. So macroscopic systems can consistently be treated as probabilistic mixture of states with different --- but definite --- values of those macroscopic degrees of freedom.

Note that nothing in this analysis relies on the environment being in any way discarded, except in the pragmatic sense that we're not terribly interested in what it's doing. The total system continues to evolve unitarily, but by virtue of the particular form of that unitary dynamics, it can consistently be given a probabilistic interpretation with respect to its macroscopic degrees of freedom.

Note also that it is the \emph{dynamical} aspects of decoherence that are important here (a point also stressed by Zurek\citeyear{zureksieve,zurekchaos}). The rapid suppression of the off-diagonal elements of the density operator, which is usually taken as the signature of decoherence, is significant not in itself but because it implies that the dynamics of the \emph{diagonal} elements must be probabilistically interpretable. (The decoherent-histories framework makes this more explicit: in that framework, the possibility of interpreting the system's evolution probabilistically is definitional of decoherence.)

\section{Decoherence and the measurement problem}

Decoherence, then, explains why the measurement problem is a philosophical rather than a practical problem. Given the ubiquity of decoherence, the strategy of applying the probability interpretation of the state to decohered systems (and to the basis with respect to which the decoherence occurs) will not actually give us contradictory predictions --- at least, not to the levels of accuracy which we have any hope of probing empirically. Those readers inclined towards the so-called ``shut up and calculate interpretation'', therefore, can stop reading now. For the rest, we should now return to the previous taxonomy of strategies for solving, or dissolving, the problem, and see how they fare in the light of decoherence.

We begin with the strategies for \emph{solving} the problem: the dynamical-collapse and hidden-variable strategies.  Recall that dynamical-collapse strategies effectively turn the shift from a physical to a probability reading of the quantum state into a dynamical process, while hidden-variable strategies effectively hold on to a physical reading of the quantum state and add additional dynamical variables over which the probabilities are defined.

At first sight, both strategies are made straightforward by decoherence. We can fairly straightforwardly specify a dynamical collapse theory by stipulating that superpositions of states in the decoherence-preferred basis spontaneously collapse, with collapse probabilities given by the probability rule; we can fairly straightforwardly specify a hidden-variable theory by treating the variable that labels the decoherence-preferred basis as a hidden variable and stipulating that the probability distribution over that variable is given by the probability rule. Neither of these recipes completely specifies the theory in question (how quickly does collapse occur? what dynamics govern the hidden variables, and does it ensure that the probabilities remain in line with the quantum predictions after the initial stipulation?) but --- it might seem --- the hard work has been done.

Unfortunately, things are not so simple, for a straightforward reason: \emph{decoherence is not a precisely defined process}. We can see this in several related ways:
\begin{enumerate}
\item Decoherence suppresses interference only approximately. Interference between alternative ``possibilities'' is far too low to be experimentally detectable --- which is another way to say that to within the limits of experimental accuracy, the probability interpretation gives consistent predictions --- but it is not zero.
\item The basis picked out by decoherence is itself only approximately given. In the standard examples, for instance, we might specify it as ``a wave-packet, not too localised in position or momentum''. But ``not too localised'' is scarcely a precise criterion; nor is there anything particularly privileged about the Gaussians usually used for wave-packets, save their mathematical convenience. Criteria like Zurek's ``predictability sieve'' \cite{zureksieve} can be used to pick a particular choice of basis, but these have more a pragmatic than a fundamental character.
\item The analysis of decoherence I gave above relies on a certain decomposition of the total Hilbert space into ``system'' and ``environment'' degrees of freedom. Varying that decomposition will vary the probability rules defined by the decoherence process. Granted, any remotely sensible decomposition can be expected to give essentially similar results, but ``remotely sensible'' and ``essentially similar'' are not really the kind of terms we'd like to see in the specification of a fundamental theory. But Dowker and Kent~\citeyear{dowkerkent} have provided strong arguments (in the rather different mathematical framework of decoherent histories) that the \emph{mere} criterion that the probability calculus applies to a given decoherence basis, divorced from other considerations, is overwhelmingly too weak to pick out a unique basis, even approximately.
\end{enumerate}

These observations all stem from the same feature of decoherence: that it is a high-level, dynamical process, dependent on details of the dynamics of the world and even on contingent features of a given region of the world. It is not the kind of thing that can be captured in the mathematical language of microscopic physics. 

Note that this is not to say that decoherence is not a real, objective process. Science in general, and physics in particular, is absolutely replete with real, objective processes that occur because of high-level dynamical effects. Pretty much any regularity in chemistry, biology, psychology, \ldots has to have this character, in fact. But \emph{emergent} processes like this don't have a place in the axioms of fundamental physics, precisely because they emerge from those axioms themselves.\footnote{Emergence has a long and tangled history in philosophy of science; see, \egc, Butterfield~\citeyear{butterfieldemergence1,butterfieldemergence2}, and references therein, for details. In physics, perhaps the most influential discussion of recent years has been \citeN{andersonmoredifferent}.}

An analogy: it has been well understood for many years that particles --- whether in relativistic quantum field theory, or in condensed matter physics --- are themselves not fundamental.  The particle spectrum of a quantum field theory is determined by the dynamical features of the theory, and in general it is determined only approximately, and in a way that varies according to the contingent features of the regime which interests us. (In some situations, we analyse quantum chromodynamics in terms of quarks; in others, in terms of protons and neutrons; not only the masses and charges of the particles, but which particles we use in the first place, vary according to the energy levels at which the theory is analysed.)

Adrian Kent puts it rather well (in a slightly different context): 
\begin{quote}
It's certainly true that phase information loss is a dynamical process
which needs no axiomatic formulation.  However, this is irrelevant to
our very simple point: no preferred basis can arise, from the dynamics
or from anything else, unless some basis selection rule is given.  Of
course, [one] can attempt to frame
such a rule in terms of a dynamical quantity - for example, some measure
of phase information loss.  But an explicit, precise rule is needed. \cite{kent}
\end{quote}

But if decoherence cannot be used to \emph{define} a set of hidden variables, or a collapse law, nonethleless it serves to \emph{constrain} both concepts. For if the collapse rule does not cause collapse onto a basis that is approximately decohered, it will fail to reproduce the experimental predictions of quantum mechanics; likewise, if a probability function over the values of the hidden variables does not determine a probability function over the decoherence-selected basis, the hidden variables will not allow us to recover the empirical predictions of quantum theory.

In the case of \emph{nonrelativistic} quantum theory, this is unprcoblematic. The decoherence-preferred basis is basically a  coarse-graining of the position basis, so a collapse rule which collapses the system onto wavepackets fairly concentrated around a particular centre-of-mass position, or a choice of position as the hidden variable, will do nicely. And indeed, we find that the main examples of non-relativistic collapse and hidden-variable theories --- the GRW theory and Bohm's theory --- do indeed select position in this way.

It is crucial to note what makes this possible. Position has a dual role in non-relativistic quantum theory: it is at one and the same time (a) one of the fundamental microphysical variables in terms of which the theory is defined, and (b) such that a coarse-grained version of it is preferred by the high-level, dynamical, emergent process of decoherence. As such, it is possible to formulate modifications or supplements to non-relativistic quantum theory that are both precisely defined in terms of the \emph{micro}physical variables used to formulate quantum mechanics, and appropriately aligned with the \emph{macro}physical variables picked out by decoherence.\footnote{Incidentally, this is what makes testing dynamical collapse theories so difficult: the fact that they must succeed in reproducing the empirical predictions of quanutum theory in normal circumstances pretty much guarantees that it will be very difficult to distinguish genuine collapse from mere decoherence. (Difficult, but not impossible; see Stamp, this volume.)}

Unhappily for modificatory strategies, there does not appear to be a variable in extant relativistic quantum theory --- in QED, say, or in the Standard Model --- that manages to play this dual role. In the fermionic sector, decoherence seems to prefer states of definite particle number --- and, as we have already seen, ``particle number'' is itself a dynamically emergent concept, determined by the complex details of renormalisation theory and dependent, to a considerable degree, on the energy levels at which we wish to study the systems of interest to us. In the bosonic sector --- at least where electromagnetism is concerned --- decoherence seems instead to select out coherent states \cite{zurekqft}, but these are coherent states defined with respect to the effective field operators governing the system at low energies. In neither case is there any remotely straightforward definition of the decoherence-preferred basis in terms of the quantities used to formulate the quantum field theory at the microphysical level. Indeed, it is a commonplace of renormalisation theory\footnote{See \citeN{peskinschroeder}, or any other modern textbook on quantum field theory, for technical details and references.} that these variables are largely hidden from view at the level of observable phenomena.

For this reason, I suspect  --- if it is accepted that modifications of quantum theory should not themselves be stated in a way which makes essential reference to dynamically emergent and high-level features of the theory --- that the prospects of solving the measurement problem in the relativistic domain by modifying quantum theory are dim. It is notable that, to my knowledge, there is no dynamical-collapse theory even purporting to be applicable to relativistic quantum theory in the presence of interactions,\footnote{The nearest thing to such a theory is Tumulka's theory \cite{tumulka}, which is explicitly formulated for a multi-particle theory, on the assumption that there are no interactions. It was, of course, precisely the need to incorporate interactions that drove the pioneers of relativistic quantum mechanics to field theory.} and those hidden-variable theories 
that have been proposed in the relativistic domain\footnote{The main examples are Durr \emph{et al}~\citeyear{goldsteinqft03,goldsteinqft04}, who take the hidden variables to be particle positions; Struyve and Westman~\citeyear{struyvewestman07}, who take them to be bosonic field configurations, and Colin and Struyve~\citeyear{colinstruyve}, who take them to be local fermion-number densities.} are largely silent about renormalisation.

Turning to the strategies for \emph{dissolving} the measurement problem, recall that there are again two: treat the wavefunction as always probabilistic, or treat it as always physical. The former strategy makes essentially no contact with decoherence: the point of decoherence (as I have presented it here) is to give an account of when a probabilistic reading of the wavefunction is consistent, but the probabilistic strategy treats it as \emph{always} consistent. It does so, in general, by retreating from any attempt to interpret the probabilities as probabilities of anything except measurement outcomes.

This strategy, in effect, is a retreat to the idea that measurement is a primitive. Insofar as that makes sense, it suffices to resolve the puzzles of interference without any concern about decoherence (as it would have to: after all, it has to explain why the probabilistic reading is \emph{always} possible, even in situations where decoherence is negligible). But, as I have alluded to earlier, it does \emph{not} make sense, so far as I can see, partly on philosophical grounds but largely on the straightforward grounds that experimental physicists, and theorists who study experiment, cannot treat measurement as primitive but invariably fall back on the need to analyse it, using quantum theory itself.

To be fair, this objection has received at least some attention from advocates of the strategy (for recent examples, see \citeN{peres} and \citeN{fuchsinformation}).\footnote{And, as I mentioned earlier, it is not \emph{universally} accepted that only a primitivist view of measurement can justify the strategy.} I leave it to readers to judge for themselves whether these responses really do justice to physical practice. It is worth pointing out, though, that in general its advocates tend to work in quantum information, a field whose \emph{raison d'etre} is to abstract away the messy details of quantum-mechanical processes and look at their abstract structure. This strategy has been remarkably successful, yielding deep insights about quantum theory that would not have come easily if we had kept the messy details in play; for all that, it's possible to worry that some quantum-information approaches to the measurement problem mistake the map for the territory.

The final strategy is Everett's: treat the quantum state as always giving the physical state of the system. If decoherence can contribute directly to any (dis)solution to the measurement problem, it is here. For we have seen that decoherence is an emergent process; what it tells us, interpreted as Everett suggests, is that even if the Universe is \emph{fundamentally} a unitarily evolving whole, at the \emph{emergent} level it has the structure of a probability distribution over states each of which describes approximately classical goings on. There is no mechanism by which one of those states is preferred (is `actual', if you like) and the others are mere possibilities: at the fundamental level, all are components in a single, unitarily-evolving state, and no one is preferred over another. (Any such mechanism would amount to a dynamical-collapse theory, as discussed previously.) But at the emergent level, each term evolves independently of the others; furthermore, their mod-squared amplitudes behave as if they were probabilities.

Does this mean that decoherence suffices to solve the measurement problem, provided that we understand the quantum state as a physical state as Everett proposed? The answer turns on two problems:
\begin{description}
\item[1. (The ontological problem):] For something to \emph{be} a collection of quasi-classical worlds, does it suffice for it to have the \emph{structure} of a collection of quasi-classical worlds --- or is more needed?
\item[2. (The probability problem):] For something to \emph{be} a probability measure over a set of quasi-classical worlds, does it suffice for it to have the  \emph{structure} of a probability measure over a collection of quasi-classical worlds --- or is more needed?
\end{description}

Space does not permit an extensive engagement with these questions. My own view\footnote{Developed in full in \citeN{wallacebook}; for earlier versions, see Wallace~\citeyear{wallacestructure,wallaceFAPP} in the first case, and Wallace~\citeyear{wallaceprobdec,wallaceprobformal} in the second. For further discussion --- on both sides --- see Saunders \emph{et al}~\citeyear{saundersetal} and references therein.} is that neither problem is truly problematic. But it is notable that both problems are essentially philosophical in nature: in the light of decoherence, if an Everettian solution to the measurement problem is to be  rejected then it will have to be for subtle philosophical reasons rather than any structural deficiency in quantum theory.

\section{Conclusion}

In twenty-first-century physics, the ``measurement problem'' is best understood, not as an illegitimate intrusion of a primitive ``measurement'' postulate into physics, but as a conceptual incoherence in our interpretation of quantum states: it seems impossible to understand the macroscopic predictions of quantum mechanics without interpreting the state probabilistically, yet because of interference, quantum states cannot systematically be thought of as probability distributions over physical states of affairs. We can attempt to resolve that incoherence either by philosophical methods (thinking hard about how to understand quantum states so as to come up with a non-incoherent way) or by modifying the physics (replacing quantum mechanics with some new theory that doesn't even prima facie lead to the conceptual incoherence).

Decoherence explains why it is that quantum theory nonetheless works in practice: it explains why interference does not, in practice spoil the probabilistic interpretation at the macro level. But because decoherence is an emergent, high-level, approximately-defined, dynamical process, there is no hope of incorporating it into any modification of quantum theory at the fundamental level. Decoherence does, however, act as a significant constraint on such modifications ---a constraint which, in the case of relativistic quantum field theory, is likely to be exceedingly hard to satisfy.

Decoherence could, however, play a role in a solution to the measurement problem which leaves the equations of quantum theory alone and treats the objective macroscopic reality we see around us as itself an emergent phenomenon. Such a strategy is committed to the claim that, at the fundamental level, the quantum state continues to describe the physical state of the world: it is, therefore, ultimately Everett's strategy. Decoherence finds its natural role in the measurement problem as the process which explains why quantum mechanics, interpreted as Everett advocates, can be fundamentally deterministic and non-classical, but emergently classical. It does not, however, in any way blunt the metaphysically shocking aspect of Everett's proposal: no one quasi-classical branch is singled out as real; all are equally part of the underlying quantum reality.

\section*{Acknowledgements}

This paper has drawn on many discussions with Simon Saunders. An earlier version was presented at the fourteenth Seven Pines symposium, generously supported by Lee Gohlike.

\end{document}